# X-ray discovery of a dwarf-galaxy galaxy collision

Gordon P. Garmire[1]

[1] Huntingdon Institute for X-ray Astronomy, LLC
10677 Franks Road, Huntingdon, PA 16652
e-mail address: g2p3g4@gmail.com

## Abstract

We report the discovery of a probable dwarf galaxy colliding with NGC 1232. This collision is visible only in the X-ray spectral band, and it is creating a region of shocked gas with a temperature of 5.8 MK covering an impact area 7.25 kpc in diameter. The X-ray luminosity is $3.7 \times 10^{38}$ ergs s$^{-1}$. The long lifetime of this gas against radiative and adiabatic cooling should permit the use of the luminous afterglow from such collisions to be used as a way of estimating their importance in galaxy evolution.

## 1. Introduction

The SAB(rc)c galaxy NGC 1232 is a nearly face-on spiral covering about 6.7' by 7.8' on the sky at a distance of 18.7 Mpc[1] with low foreground X-ray absorption, making it an ideal galaxy for study by the *Chandra X-ray Observatory* ACIS-I[2] array. The ACIS-I array encompasses an area of 16.9' by 16.9'. There is a smaller galaxy, NGC 1232A, that lies 4.07' to the southeast of NGC 1232, having a recessional velocity of 6599 km/s at an estimated distance of 2.4 Mpc farther away than NGC 1232 is. It is not likely to be currently interacting with NGC 1232. Both galaxies may be associated with the Eridanus group of galaxies, sometimes referred to as a galaxy cluster or group of groups (Willmer, et al. 1989), but at a projected distance of 2.2 Mpc from the center of the low mass cluster, neither are likely to be bound to the cluster. At this large projected distance there is no X-ray emission expected from the cluster or group based on observations of other clusters and groups (Miller et al. 2012). The density of any gas associated with the cluster must be extremely low ($<10^{-4}$ cm$^{-3}$) and incapable of affecting the gas in an individual large spiral galaxy.

No previous *Chandra* observation of NGC 1232 has been made. An initial observation of 100ks was proposed, that reaches a limiting luminosity of $2 \times 10^{37}$ ergs/s and should provide over 100 X-ray sources for study if the source population is typical (Prestwich et al 2009). The purpose of the initial study was to classify the X-ray sources based on X-ray colors, to search for time variability and to study the diffuse X-ray emission. The source studies are still ongoing, but in examining the diffuse emission an unusual pattern of emission was discovered, that is presented in the following paper.

It appears that an enormous cloud of gas is colliding with the nearly face-on spiral galaxy NGC1232. The impact is creating a region of shocked hot gas with a temperature of about 5.8 MK. The smoothed image of the collision is shown in Figure 1a and 1b and reveals a cloud with a cometary appearance sweeping across the galaxy and possibly colliding with the disk. The center of the collision is about 4.3 kpc from the center of NGC 1232. The collision region is very large, with a diameter for the most intense X-ray emission covering about 80" or 7.25kpc for an assumed distance of 18.7 Mpc. This size suggests that a dwarf galaxy may be involved in a collision with NGC 1232. The long "tail" of the X-ray emission implies that the dwarf galaxy is passing through the hot halo of NGC 1232, creating a shock heated region in the halo. Such collisions visible only in X-rays may represent an increase in the estimated growth rate of large galaxies from galaxy mergers. The shape of the hot gas is quite different from the diffuse emission in other spirals, making it unlikely that this emission is due to star formation regions that often show soft X-ray emission, but with much smaller sizes (typically < 100pc) and which usually follow the pattern of the spiral arms.

---

[1]This distance is taken from the NASA/IPAC Extragalactic Database (NED) based on 9 estimates in the literature that has a standard error of 2.6 Mpc
[2]Advanced Charged Coupled Device Imaging Spectrometer (ACIS) Imaging array.

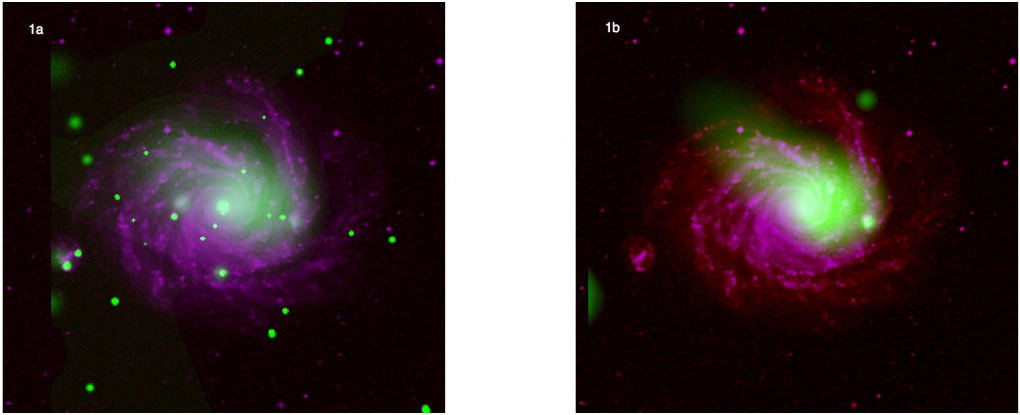

Figure 1a. The cloud superimposed on NGC 1232 with the X-ray point sources retained. Point sources are either white or green depending upon whether the source is superimposed on the galaxy or not. Purple is the optical image of the galaxy NGC1232 and its environs. The X-ray image has been smoothed using the tool *csmooth*. Figure 1b. The cloud superimposed on NGC 1232 with the point X-ray sources removed and the X-ray image smoothed using *csmooth*. The green sliver of an image to the lower left is a cluster of galaxies seen as an X-ray source at the edge of the frame. Without the point sources, the fainter X-ray emission can be enhanced to bring out more details.

## 2. Observations and Data Analysis

The observations reported here of NGC 1232 were made by the *Chandra X-ray Observatory* starting in 2008 and ending in 2010. The exact times are listed in Table 1.

Table 1
The Observations

| Observation Start Time | Exposure time(s) | ObsID |
|---|---|---|
| 2008 Nov 03, 20:18:03 | 47012 | 10720 |
| 2008 Nov 05, 22:57:50 | 52229 | 10798 |
| 2010 Sep 29, 17:42:26 | 48416 | 12153 |

The observations were made using the Advanced CCD Imaging Spectrometer imaging array (ACIS-I; Garmire et al. 2003). The ACIS-I utilizes four front-illuminated CCDs arranged in a square pattern with small gaps between the CCDs. Each CCD utilizes 24 micron square pixels (0.492" on the sky) in a 1024 x 1024 array. The field of view of the ACIS-I array is 16'.9 x 16'.9. which is much larger than the 8' size of the galaxy. The CCDs were operated in the Very Faint mode, which telemeters the 5 x 5 pixel region around each event. This mode improves background rejection. The focal-plane temperature was held at -120 C during these observations, assuring stable operation of the CCDs.

The light curves of each observation were examined using the software program Event Browser in the Tools for ACIS Real-time Analysis (TARA; Broos et al. 2000) for possible background flaring that is sometimes observed during an exposure. No flares outside of statistics for a constant background were detected, making all of the data suitable for analysis. The analysis of the data employed the *Chandra* Interactive Analysis of Observations (CIAO) tools version 4.2. The point sources were identified using the CIAO tool *wavdetect*. The source regions were constructed using data from the telescope point spread function (PSF) and excluded from the image. Pixels inside the PSF were filled based on the local image intensity using the CIAO tool *dmfilth*. The DIST method was used to fill in the background, since the count rates were not sufficiently high to use the POISSON method. The image was binned by a factor of four to better search for extended diffuse emission. The image was filtered into several energy bands to search for diffuse emission at different temperatures. A cometary cloud feature, most apparent in the energy range 500 eV to 1000 eV, is shown in Figure 1 after running the CIAO tool *csmooth* over the data setting the minimum and maximum signal-to-noise between 3 and 5 for the FFT smoothing algorithm.

## 3. Discussion

The cloud probably represents a very large collisional event between a dwarf galaxy and the large spiral galaxy NGC 1232. An analysis of the X-ray spectrum of the event, which represents 274 photons in the extraction region above background, using the tool XSPEC (Arnaud 1996) gives a temperature of 5.8 MK assuming a Nonequilibrium Ionization (NEI) plasma model for the emission with 0.3 Solar cosmic abundance. The result is not sensitive to the abundance. Table 2 summarizes the spectral parameters of the X-ray emission. For the NEI model the $n\tau$ parameter is $2.2 \times 10^{11}$ s cm$^{-3}$ indicating that the plasma is not in equilibrium, except the errors are large enough to include equilibrium at the upper end of the range.

Table 2
Spectral fit parameters

| Model | $N_H$(x $10^{22}$cm$^{-2}$) | kT(keV) | Log($n_e\tau$(s cm$^3$)) | EI(cm$^{-3}$) | $F_{unabs}$(ergs cm$^{-2}$ s$^{-1}$) | $\chi^2_{dof}$ |
|---|---|---|---|---|---|---|
| APEC | 0.02[a] | $0.47^{+0.18}_{-0.17}$ | | $6.5 \times 10^{60}$ | $8.4^{+1.17}_{-2.2} \times 10^{-15}$ | 0.44 |
| NEI | 0.02[a] | $0.50^{+0.09}_{-0.08}$ | $11.34^{+2.36}_{-0.49}$ | $7.4 \times 10^{60}$ | $9.4^{+1.8}_{-1.8} \times 10^{-15}$ | 0.40 |

[a]The value for the absorption in our galaxy. There were insufficient counts from the cloud to determine the $N_H$ accurately so it was fixed at the galactic value for the fit to the temperature and the flux. The Flux is between the energies 0.5 – 2.0 keV. EI, the emission integral assumes the spherical cloud geometry,

The region of the cloud enclosed in the red contour was selected for the spectral fit and is shown in Figure 2, where several contours are displayed. This region contains the bulk of the counts and is a factor of greater than two above the background level. The shape of the cloud was visible in all three observations and Figure 2 represents the combined data from all of these. The CIAO tool *merge_all* was used to combine the data about a common RA and Dec.

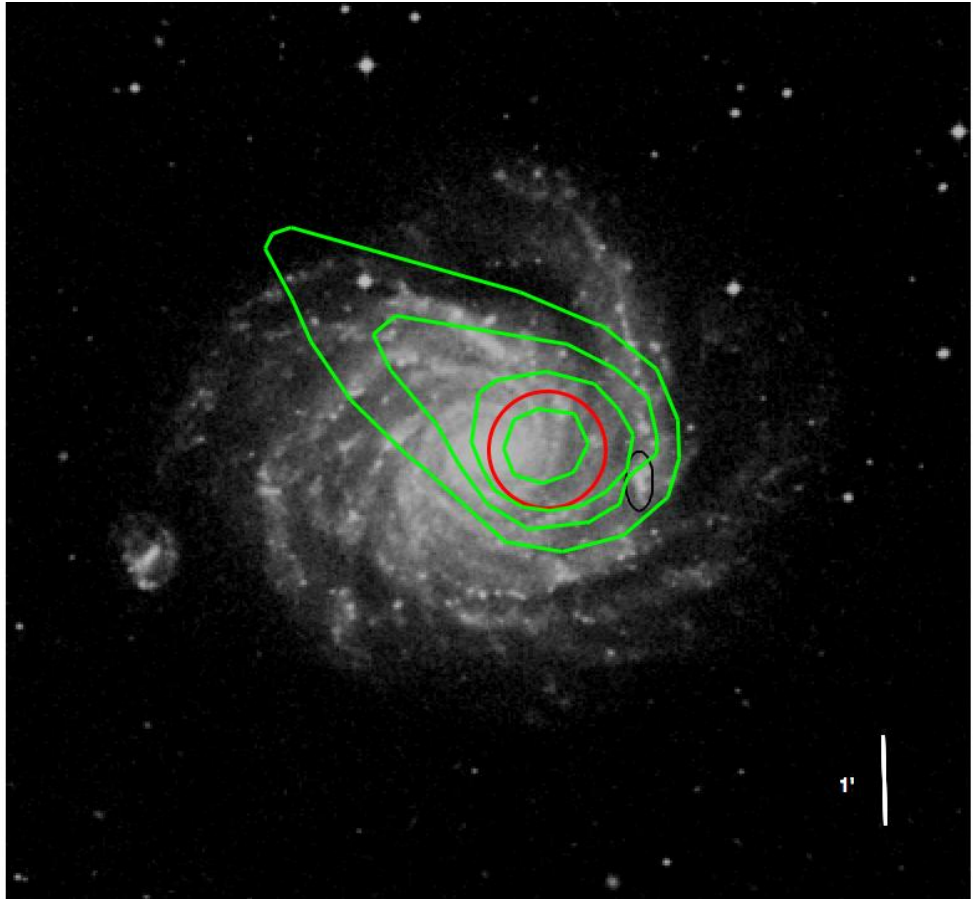

Figure 2. The galaxy NGC 1232 with contours of X-ray emission overlaid (from the Digitized Sky Survey, 1994DSS…1…0000). The red contour is the extraction region for the spectral analysis. The inner green contour is set at a level of $5.7 \times 10^{-15}$ ergs cm$^{-2}$ s$^{-1}$ square arc minute$^{-1}$. Proceeding outward the next green contour is set at $4.0 \times 10^{-15}$, the next at $2.6 \times 10^{-15}$ and the outer contour at $1.0 \times 10^{-15}$. The black contour is the H II region that is seen to emit X-rays at a detectable level.

The neutral hydrogen in our galaxy in the direction of NGC 1232 is estimated to be $2 \times 10^{20}$ atoms cm$^{-2}$ according to Colden[3] of the CIAO Proposal Tools and was fixed for the fits. There could be some additional absorption in the NGC 1232 environs, but there is no reliable way to estimate this and the fit to the spectrum is rather insensitive to the exact value in any case. The NEI spectrum obtained from the XSPEC tool is shown in Figure 3 with the parameters listed in Table 2 and the abundance set at 0.3 of the Solar value. This value is rather arbitrary, but the parameters are not very sensitive to variations in the abundance.

[3]http://cxc.harvard.edu/toolkit/jsp

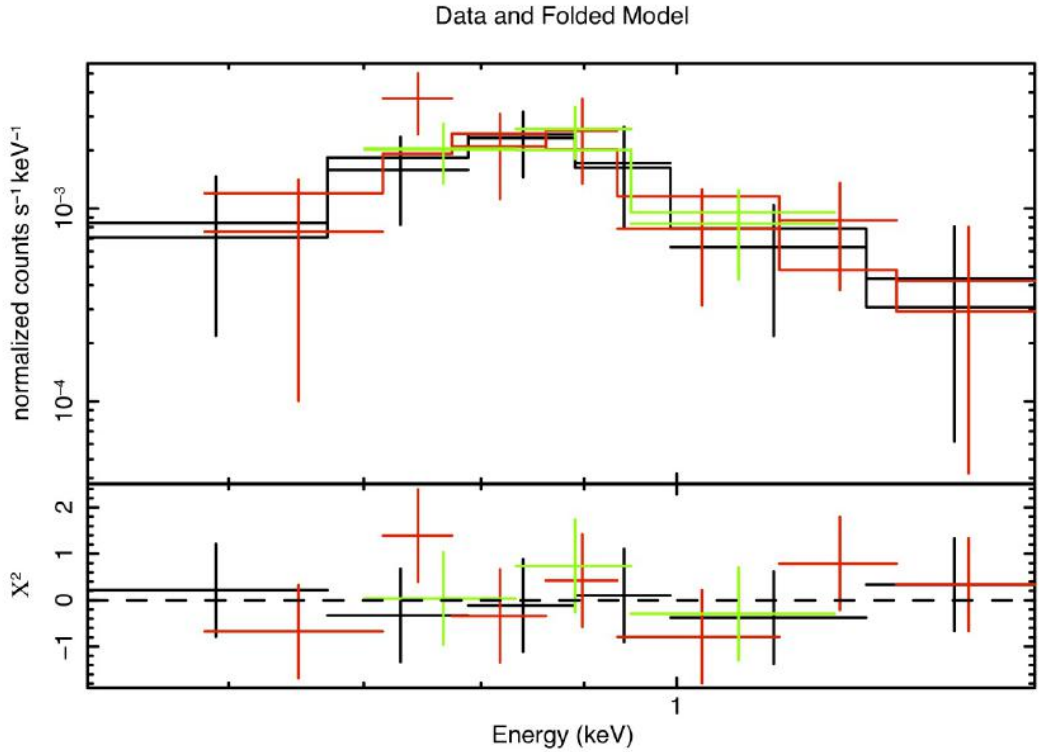

Figure 3. The spectrum of the brightest portion of the cloud shown in Figure 2 fitted with an NEI model. The three colors of the data points correspond to the three observations listed in Table 1; black from ObsID 10720, Red from ObsID 10798 and green from ObsID 12153. The fit is a simultaneous fit to all three data sets. The parameters are listed in Table 2.

The three dimensional geometry of the cloud is indeterminate, so that several possible configurations will be considered. First consider the case where the bright forward end of the cloud is approximately spherical. Taking the distance to NGC 1232 as 18.7 Mpc, the angular diameter of 80" implies a linear diameter of 7.25kpc. This large diameter implies a galactic sized event is in progress. If we assume that the emission is primarily thermal bremsstrahlung from a hot plasma as indicated by the spectral fit to the X-ray spectrum, then we can make an estimate of the amount of mass and energy involved in the hot plasma. The volume of the core of the cloud assuming a spherical geometry is $5.4 \times 10^{66}$ cm$^3$ and the luminosity is $3.7 \times 10^{38}$ ergs s$^{-1}$. Assuming that the density of the cloud is uniform leads to an upper limit to the electron density since the emission scales as $N_e^2$ thereby generating more emission from high density regions and requiring less mass to create the same luminosity. This uniform density is $1.4 \times 10^{-3}$ cm$^{-3}$ which implies a total mass of hot gas for the cloud of $3 \times 10^6$ Solar masses and a thermal energy of $1.8 \times 10^{55}$ ergs. This mass is comparable to several dwarf galaxy masses in the Local Group of dwarf galaxies (M. Mateo 1998)

If we assume that the cloud is interacting only through the collision with the disk of NGC 1232 in the form of a shock front, then we might take a more disk shaped region. The thickness of the disk is the most difficult parameter to estimate, but the reverse shock as in the case of a supernova remnant could be a parsec or more in thickness. If we take a parsec as a lower limit for the emission thickness, then the volume of the emission is $1.1 \times 10^{63}$ cm$^3$. The luminosity is the same and the electron density is $9.7 \times 10^{-2}$ cm$^{-3}$. At this density the thermal energy of the cloud is 2.5 x 10$^{53}$ ergs and the mass of hot gas is 4.3 x 10$^4$ solar masses. These masses are consistent with the size of dwarf galaxies (Sargent & Lo, 1985; Mateo, 1996), although the mass of stars in the cloud is unknown, but is probably at least an order of magnitude more than the hot gas. The lower mass for the disk shaped model is low for a dwarf galaxy, more like a globular star cluster, but the physical size of the distribution is much greater than that of a globular star cluster.

For either assumed geometry, the thermal energy of this event is quite large, being three to five orders of magnitude greater than a typical supernova remnant. The hot gas has a long radiative cooling time (~10 Myr to 620Myr) depending upon the geometry, implying that the disk of the galaxy should show an enhanced region of emission for at least one galactic rotation. The collision timescale is of the order of 50 Myr assuming that the dwarf galaxy is moving at about the escape velocity of NGC 1232. The adiabatic cooling time is likely to be shorter than the radiative cooling time, but without knowledge of the external pressure on the cloud, it is rather uncertain. The frequency of these collisions with dwarf galaxies is difficult to estimate, but it is unlikely to be any more frequent than once every Gyr otherwise most galaxies would show signs of collisions with dwarf galaxies. They do not show large, extended hot regions (see Tyler et al. 2004) except along arms in spiral galaxies. This being the case, supernovae and massive stellar winds still dominate the heating of the interstellar medium at this stage of galactic evolution. At earlier times, these collisions were probably much more frequent and contributed significantly to galaxy growth. Nevertheless, examining galaxies for large regions of hot gas may be a way to estimate the frequency of collisions with dwarf galaxies and how much such collisions contribute to galaxy growth and evolution in the current epoch.

This appears to be the first case of a dwarf galaxy-large galaxy collision that is visible only in the X-ray band. We have examined radio (Zee & Bryant, 1999 and Condon 1987), IR (Jarrett et al. 2003) and Hα (van Zee et al. 1998 and Garcia-Gomez & Athanassoula 1991) images and see no evidence of a disturbance in these bands associated with this collision. The mass of the dwarf galaxy is four orders of magnitude or more, less than the mass of NGC1232 making it unlikely that any substantial disturbance in the structure of NGC1232 would be detectable. In Figure 2 there is a collection of very bright stars at the head of the second-most outer contour. This region is defined by a black ellipse. These bright stars are associated with an X-ray enhancement that is extended (~8″) with an X-ray peak surface brightness of 5.2 x 10$^{-15}$ ergs cm$^{-2}$s$^{-1}$ per square arc minute. This corresponds to a luminosity of 4.6 x 10$^{37}$ergs s$^{-1}$ for an assumed distance of 18.7Mpc. This luminosity is quite high for a large H II region. Even 30 Doradus, one of the largest with a diameter of 250 pc and the most luminous H II regions in the LMC only has a luminosity in X-rays of 3.1 x 10$^{37}$ ergs s$^{-1}$ (Townsley et al. 2011). The implied size of the region is also very large for an HII region being about 1 kpc in diameter. This is much larger than any known HII region. It is also larger than any supernova remnant. Perhaps it is a large bubble carved out of the ISM by multiple supernovae and hot stellar winds. In our galaxy the Cygnus Superbubble approaches the scale of this feature, but falls short by an order of magnitude in luminosity and energy content (Cash et al. 1980). The HII region is at the leading edge of the cloud, which could indicate that the shock wave has triggered star formation in this region.

High velocity H I clouds have been observed interacting with our galactic halo, much as this galaxy collision is interacting with NGC 1232's halo, to produce X-ray emission (Kerp et al. 1996) but on a much smaller scale. Bruns, Kerp and Pagels, 2001 observed Compact High Velocity Cloud (CHVC) 125+41-207 with a cometary appearance at a distance of about 130 kpcs, well out of the galactic plane. This appears to be a miniature version of the cloud colliding with NGC1232, but visible in the radio instead of the X-ray band. Modeling of collisions with HVCs by Tenorio-Tagle et al. 1987 indicates that temperatures of over 1 MK are attained even for collisions with relatively small clouds at modest velocities.

An alternative interpretation of this event is that it represents a very large region of hot gas in NGC 1232 that is off-set from the nuclear region. Several other galaxies exhibit hot gas extending along the arms of the galaxy such as M101 (Kuntz & Snowden, 2010) and NGC 2903 (Yukita et al. 2012). Also see Tyler et al. 2004, for a survey of 12 nearby spiral galaxies that are studied. It should be noted that none of these galaxies exhibit such a large region of X-ray emission. The X-ray emission in these galaxies is attributed to star formation resulting in subsequent supernova activity and massive hot stellar winds that have diffused out into the arm regions. Under this assumption, the thickness of the hot emitting region could be as much as 1 kpc, in which case the emitting volume is about 1.7 x 10$^{66}$ cm$^3$, the average density, assuming uniform density, is 3 x10$^{-3}$ cm$^{-3}$, the energy content is 8 x 10$^{54}$ergs and the mass of hot gas is 1.3 x 10$^6$ solar masses. The temperature of the hot gas in other galaxies is typically 2 – 6 MK, not different from the hot gas in NGC1232. Such a large amount of energy would require nearly $10^5$ SNR and/or O-stars to power this hot gas. The radiative cooling time for this gas is of the order of 280 Myr, however adiabatic expansion could provide a more rapid means of cooling the gas. With such a long cooling time it is possible to provide the energy to maintain the hot plasma, but differential rotation in the galaxy should smear out the plasma making the distribution more cylindrically symmetric about the center of the galaxy. The southern half of NGC 1232 shows no detectable X-ray emission above a 2σ upper limit of 3 x $10^{-15}$ergs cm$^{-2}$ s$^{-1}$. There is no evidence in the radio or IR bands for absorption or excess gas that might mask X-ray emission from the southern half of NGC 1232. Even including the X-ray cloud, the total X-ray luminosity is about average for a spiral galaxy based on the Mid IR emission of 1.7 x $10^{-11}$ ergs cm$^{-2}$ s$^{-1}$ (Sanders et al. 2003 and Tyler et al. 2004). The range of spiral galaxy IR luminosities is a factor of five in Figure 2 of Tyler et al. 2004 at the flux level of the cloud, but without the cloud, the remaining X-ray diffuse luminosity of NGC 1232 falls lower than any of the galaxies in the studies of Tyler et al. 2004, although the 2σ upper limit puts NGC 1232 at the lower bound of the other spirals in the study.

## 4. Conclusions

The *Chandra X-ray Observatory* observations of the galaxy NGC1232 have revealed a cometary shaped cloud of hot (5.8MK) gas superimposed on the optical galaxy (Fig. 1). The origin of this hot cloud is presumed to result from the passage of a dwarf galaxy through the halo of NGC 1232 and possibly by the collision of the dwarf galaxy with the gas disk of NGC 1232. Based on the X-ray spectral measurements and X-ray intensity some rough parameters of the hot gas may be derived. The result is that the hot cloud is emitting at a level of 3.7 x $10^{38}$ ergs s$^{-1}$ and has a mass of hot gas ranging between 4.3 x$10^4$ and 3 x$10^6$ solar masses depending upon the assumption of spherical or pancake geometry for the cloud. The masses of dwarf galaxies cover a wide range both of the gas content and the stellar content, but the parameters derived here for this cloud are consistent with the range of masses for dwarf galaxies. Such collisions are probably rare in the current epoch, because the cooling residue of these collisions, which should be visible in a substantial fraction of galaxies, is not observed. The alternative explanation that this feature is the result of a large number of supernovae and hot stellar winds is not supported by any optical, IR or radio features that are only on one side of the galaxy as is this cometary cloud.

The author would like to thank Audrey Garmire for her contributions to the data analysis, a careful reading of the manuscript and thoughtful discussions on the scientific interpretation.

This work was supported In part by the ACIS Instrument Team contract SV4-74018 (PI: G. P. Garmire) and by the contract SV2-82024 issued by the *Chandra* X-ray Center, which is operated by the Smithsonian Astrophysical Observatory for and on behalf of NASA under contract NAS8-03060. This research has made use of the NASA/IPAC Extragalactic Database (NED) which is operated by the Jet Propulsion Laboratory, California Institute of Technology, under contract with the National Aeronautics and Space Administration.

References

Arnaud, K. 1996 in ASP Conference Ser. 101, Data Analysis Software and Systems V. ed. G H. Jacoby & J. Barnes (San Francisco: ASP), 17

Broos, P., et al. 2000 Users Guide for the TARA Package: Document Revision 5.8 (Universiy Park, Penn Sate University

Brüns, C., Kerp, J., & Pagels, A. 2001, "Deep HI observations of the compact high-velocity cloud HVC 125+41-207" A&A 370, L26.

Cash, W., Charles, P., Bowyer, S., Walter, F., Garmire, G., & Reigler, G. 1980, 238, ApJL 71-76.

Condon, J. J. 1987 ApJS **65**, 485-54

Garcia-Gomez, C. & Athanassoula, E. 1991 A&AS **89,** 159-184

Garmire, G. P., Bautz, M. W., Ford, P. G., Nousek, J. A., & Ricker, G. R. 2003, Proc. SPIE, 4851, 28

Jarett, T., Chester, T., Cutri, R., Schneider, S., & Huchra, J. 2003 AJ 125, 525

Kerp, J., Mack, K. –H., Egger, R., Peitz, J., Zimmer, F., Mebold, U., Burton, W. B., $ Harmann, D. 1996 A&A v312, p67-73

Kuntz, K. D. & Snowden, S. L., 2010 ApJS 188, 46-74

Mateo, M. 1998, Annu. Rev. Aston. Asttophys. 36, 455-506

Miller, Eric D.; Rykoff, Eli S.; Dupke, Renato A.; Mendes de Oliveira, Claudia; Lopes de Oliveira, Raimundo; Proctor, Robert N.; Garmire, Gordon P.; Koester, Benjamin P.; McKay, Timothy A. , 2012 ApJ 747, 94-107

Prestwich, A. H., Kilgard, R.E., Primini, F., MacDowell, J.C., & Zezas, A. 2009 ApJ 705, 1632

Sanders, D. B., Mazarella, J. M., Kim, D.-C., Surace, J. A., & Soifer, B. T. 2003 ApJ **126**, 1607-1664

Sargent, W. L. W. & Lo, K-Y. 1985, Proc. Workshop on "Star Forming Dwarf Galaxies and related Topics", ed D. Knuth, T. X. Thuan & J. T. T. Van (kim Hup Lee Printing Co. Ote, Ktd.) 253

Tenorio-Tagle, G., Franco, J., Bodenheimer, P. & Rozyczka, M. 1987 A&A 179, 219-230

Townslcy, L., et al. 2011, ApJS, 194:16

Tyler, K., Quillen, A. C., LaPage, A., & Rieke, G. H. 2004 ApJ 610, 213

Van Zee, L. & Bryant, J. 1999, AJ 118, 2172

Van Zee, L., Salzer, J. J., Haynes, M. P., O'Donoghue, A. A., & Balonek, T. J., ApJ **116**, 2805-2833

Willmer, C. N.A., Focardi, P., Nicloaci Da Costa, L., & Pellegrini, P. S., 1989 ApJ, 98, 1531-1541

Yukita, M., Swartz, D. A., Tennant, A. F., Soria, R., & Irwin, J. A., 2012 ApJ 758, 105